\documentclass[preprint2]{aastex}
\shorttitle{Streamer Instability}

\shortauthors{Chen, Li, Song, Shi, Feng, {\&} Xia}

\begin{document}

\title{Intrinsic Instability of Coronal Streamers}

\author{Y. Chen\altaffilmark{1}, X. Li\altaffilmark{2},
H. Q. Song\altaffilmark{1}, Q. Q. Shi\altaffilmark{1}, S. W.
Feng\altaffilmark{1}, AND L. D. Xia\altaffilmark{1}}
\altaffiltext{1}{School of Space Science and Physics, Shandong
University at Weihai, Weihai Shandong, China 264209;
yaochen@sdu.edu.cn} \altaffiltext{2}{Institute of Mathematics and
Physics, University of Aberystwyth, UK}

\begin{abstract}
Plasma blobs are observed to be weak density enhancements as
radially-stretched structures emerging from the cusps of quiescent
coronal streamers. In this paper, it is suggested that the
formation of blobs is a consequence of an intrinsic instability of
coronal streamers occurring at a very localized region around the
cusp. The evolutionary process of the instability, as revealed in
our calculations, can be described as follows, (1) through the
localized cusp region where the field is too weak to sustain the
confinement, plasmas expand and stretch the closed field lines
radially outwards as a result of the freezing-in effect of
plasma-magnetic field coupling; the expansion brings a strong
velocity gradient into the slow wind regime providing the free
energy necessary for the onset of the subsequent
magnetohydrodynamic instability; (2) the instability manifests
itself mainly as mixed streaming sausage-kink modes, the former
results in pinches of elongated magnetic loops to provoke
reconnections at one or multi locations to form blobs. Then, the
streamer system returns to the configuration with a lower cusp
point, subject to another cycle of the streamer instability.
Although the instability is intrinsic, it does not lead to the
loss of the closed magnetic flux, neither does it affect the
overall feature of a streamer. The main properties of the modelled
blobs, including their size, velocity profiles, density contrasts,
and even their daily occurrence rate are in line with available
observations.
\end{abstract}


\section{INTRODUCTION}
Coronal streamers are quasi-stationary large-scale structures
observed during solar eclipses or with coronagraphs, they are
believed to be the outcome of a complex magnetohydrodynamic (MHD)
coupling between the hot expanding coronal plasmas and the
large-scale confining magnetic field (e.g., Pneuman \& Kopp, 1971;
Koutchmy \& Livshits, 1992). Many streamers remain visible over a
period of up to several months. Nevertheless, various types of
activities associated with coronal streamers are observed,
including small-scale plasma blobs (Sheeley et al., 1997; Wang et
al., 1998, 2000), streamer detachments with or without in/out
pairs (Wang \& Sheeley, 2006; Sheeley \& Wang; 2007; Sheeley,
Warren \& Wang 2007), streamer blowout coronal mass ejections
(CMEs) (Howard et al. 1985; Hundhausen 1993), and others (like
streamer puffs (Bemporad et al. 2005), streamer deformations and
deflections possibly induced by nearby coronal disturbances (e.g.,
Hundhausen 1987; Sheeley, Hakala, \& Wang 2000)).

Among these activities, we are particularly interested in the
formation of plasma blobs, which were discovered with the Large
Angle Spectrometric Coronagraph (LASCO) onboard the Solar and
Heliospheric Observatory (SOHO) (Brueckner et al. 1995), first
reported by Sheeley et al. (1997). They manifest themselves as
weak density enhancements emerging from the elongated tips of
coronal streamers and confined to the narrow plasma sheet without
causing any permanent disruption or dislocation of streamers.
Since they are thought to flow passively along the bright plasma
sheet and get dynamically coupled to the ambient plasmas, blobs
are regarded as tracers of the ambient flow velocity with only
slight contribution to the total slow wind mass flux. The
estimated density inside the blobs is weakly enhanced over its
surroundings by less than, say, 10 {\%}. The sizes, when blobs are
first observed, are about 1 R$_\odot$ in the radial direction and
0.1 R$_\odot$ in the transverse direction(corresponding to about
2$^{\circ}$ in latitudinal extension). The blob speeds generally
increase from 0 - 250 km s$^{-1}$ when first observed with LASCO
C2 in the region 2 - 6 $R_\odot$ (solar radii) to about 250 - 450
km s$^{-1}$ in the region 20 -30 $R_\odot$ (Sheeley et al. 1997).
It was shown by Wang et al. (1998) that the occurrence rate of
blobs is about four per day along an edge-on streamer in the April
of 1997, the quiet period of solar activities. During times of
solar eruptions, the number, size and brightness of blobs
generally increase (Wang et al. 2000).

Theoretical modelling of coronal streamers started early in the
1970s (e.g., Pneuman \& Kopp, 1971). The classical model developed
by Pneuman \& Kopp treated the streamer as a magnetostatic
structure in a bipolar configuration with a thin axisymmetric
current sheet. The sheet extends from the tip of the closed
streamer arcades to infinity, which separates the open field lines
with opposite directions. Subsequent models presented similar
morphological picture of streamers, many of them concentrate on
the heating and acceleration of the corona and solar wind plasmas
emphasizing the steady or quasi-steady feature of the streamer
structure (e.g., Steinolfson et al., 1982; Wang et. al., 1993,
1998; Suess et al., 1999; Chen \& Hu, 2001; Hu et al., 2003a;
Suess \& Nerney, 2006). Some authors did mention that it may be
difficult to obtain an exact fully-converged steady state of the
coronal streamer. For instance, Washimi et al. (1987) found that
the modelled values of the velocity and density have a consistent
temporal variation of about 1 $\%$ in magnitude per hour; Suess et
al. (1996) demonstrated that the streamer is subject to slow and
continuing evaporation till all the closed field lines are fully
opened if heated with a volumetric heat source. Endeve et al.
(2003, 2004), on the other hand, showed that coronal streamers can
undergo eruptions if a specific heat source deposited inside the
closed arcades is added to protons rather than electrons. It
should be pointed out that the numerical resolution along the
latitudinal direction in their studies is taken to be
2.77$^{\circ}$ between adjacent grid points, which is even larger
than the latitudinal size of the blobs observed by Sheeley et al,
as introduced previously. Therefore, in these models it is
impossible to resolve the blob structures as well as their
formation process.

In Einaudi et al. (1999), the formation of blobs is simulated as
magnetic islands resulting from the nonlinear development of the
tearing mode instability along a hypothetical one dimensional
Harris neutral sheet. The sheet is assumed to be embedded in a
prescribed fluid wake, as a simplification of the realistic
configuration associated with the streamer belt above the cusp.
Refinements of this oversimplified incompressible model considered
the effects of compressibility and inhomogeneous density
distribution in relevance to the dense plasma sheet (Einaudi et
al., 2001). By further considering the two-dimensional
configuration with open and closed magnetic field lines and the
specific cusp geometry, Lapenta {\&} Knoll (2005) concluded that
the reconnections forming the plasma blobs are driven by
converging flows at the cusp region with reconnection rate mainly
determined by the driving flows instead of resistivity, allowing
their results to be extended to realistic coronal Lundquist
numbers. However, their model is based on the Cartesian geometry
with prescribed wake velocity field, and the imposed streamer
morphology is not result from the self consistent dynamical
equilibrium between the plasma expansion and magnetic confinement.
In this paper, we present our theoretical endeavor in improving
these relevant previous models for a better understanding of the
formation of streamer blobs. The improvements are three-folds.
Firstly, our calculations are conducted in the spherical geometry,
which is somewhat more realistic compared with the planar
cartesian geometry. Secondly, the streamer system in our model is
self-consistently determined by the MHD coupling between plasmas
and magnetic fields. Lastly, we take advantage of a much higher
numerical resolution to better resolve the physical processes in
the blob forming region. We show that the blobs are the
consequence of an intrinsic instability of coronal streamers
associated directly with the localized cusp region. In the
following section we reveal the details of our theoretical model
with initial and boundary conditions. The obtained solution for
the streamer solar wind-blob system is shown in section 3. A brief
summary of this article with discussion is provided in the last
part of this article.

\section{Theoretical models}

Assuming the corona to be axisymmetric as in most previous
studies, we solve the full-set two-dimensional and three-component
(2.5d) ideal MHD equations in the spherical coordinate system
($r,\theta,\varphi$). A magnetic flux function $\psi(t,r,\theta)$
is introduced to express the magnetic field
$\emph{\textbf{B}}(B_r, B_\theta, B_\varphi)$ as
$$
\emph{\textbf{B}}= \nabla\times({\psi \over
r\sin\theta}\mathbf{\hat{\varphi}})
+B_\varphi\mathbf{\hat{\varphi}}. \eqno{(1)}
$$
The deduced equations are given in Hu (2004) and are not to be
repeated here. The equations are solved numerically in a domain of
$R_\odot \le r \le 30 R_\odot$ and $0 \le \theta \le 90^{\circ}$,
which is discretized into 300 $\times$ 632 grid points. Note that
we also assume the system to be symmetric about the equator, thus
the parameters in the other half of the latitudinal domain is
derived from the obtained solution. The grid spacing increases
according to a geometric series of a common ratio 1.015 along the
radial direction from 0.005 $R_\odot$ at the solar surface to
0.437 $R_\odot$ at the top boundary. In the latitudinal direction,
we first adopt a uniform mesh between $\theta=90^{\circ}$ to
$\theta=30^{\circ}$ with grid spacing taken to be 0.1$^{\circ}$,
then we let the grid spacing increase according to a geometric
series of a common ratio of 1.134 from $\theta=30^{\circ}$ to the
pole. Such arrangement allows us to resolve the thin current
sheets above the streamer cusps and the detailed dynamics of
plasma blobs with high resolution. The multistep implicit scheme
(MIS) developed by Hu (1989) is used to solve the MHD equations.
The scheme has a relatively low numerical dissipation, which has
been employed to deal with various issues in space physics, like
the two-dimensional corona and solar wind solution driven by
Alfve\'nic turbulence (Chen {\&} Hu, 2001; Hu et al., 2003a, Li et
al. 2004, 2006 ), and the flux rope catastrophe model for CMEs (Hu
et al., 2003b; Chen et al., 2006, 2007). Since we are solving the
equations under the framework of ideal MHD, any reconnections
involved in the numerical solutions stem from numerical rather
than physical resistivity. Therefore, the low dissipative feature
of the adopted numerical scheme is critical to the issues we are
concerned.

The heating and acceleration mechanisms of the corona and solar
wind plasmas, which still remain unresolved at the present time,
are approximated by using a polytropic process with the polytropic
index taken to be $\gamma=1.05$. To start our modelling, the
initial magnetic field is taken to be a quadrupolar one with the
azimuthal component set to be zero and the magnetic flux function
given by Antiochos et al. (1999) as follows,
$$
\psi(r,\theta)={\psi_c}\{{\sin^2\theta R_\odot \over
r}+{(3+5\cos2\theta)\sin^2\theta{R_\odot}^3 \over 2r^3}\}
\eqno{(2)}
$$
where $\psi_c={\psi_0 B_p \over {10 B_0}}$, the unit of magnetic
field strength $B_0=0.833$ Gauss, the unit of magnetic flux
function $\psi_0=4.04\times 10^{13}$ Wb, and the magnetic field
strength at the pole $B_p$ is taken to be 8 Gauss. It is known
that a neutral point exists in this potential field at $r_N=\sqrt3
R_\odot$ along the equator with $\psi=0.3697 \psi_0$. The reasons
we adopt this specific topology, rather than the often used
bipolar topology for this study are given as follows. First, the
triple-streamer system is more realistic, especially when one
models the corona at high solar activity. As a matter of fact,
even at solar minimum, the corona is often observed to have a few
rather than one-single well-defined streamers. Second, this allows
us, in future development of the model, to extend the break-out
CME model originally proposed by Antiochos et al. (1999) to the
case including the effect of the solar wind. Further elaboration
concerning the relevance to the CME modelling will be given in the
last section. Third, by focusing on the dynamics of the side
streamer which is fully located inside the calculation domain, we
avoid the influence of the symmetric boundary conditions along the
equator.

In Figure 1a we show the magnetic field lines of the initial
quadrupolar field from 1 to 10 $R_\odot$, represented by the
contours of the magnetic flux function with the contour level
separated by $\psi=0.05 \psi_0$. On the solar surface, the
magnetic flux function first increases from 0 at the pole to 1.2
$\psi_0$ near the center of the side arcade, and then decreases
again to 0 at the equator. The separatrix in the initial
quadrupolar field specified by $\psi=0.3697 \psi_0$ is shown as
red curves. It is easy to see that the initial magnetic field
consists of four flux system including the central arcade system
straddling the equator, two side ones centered at
$\theta=45^{\circ}$ and $135^{\circ}$, and an overlying polar flux
system. As will be clearly illustrated in the following section,
the initial magnetic topology is subject to significant change by
the plasma-field coupling process.

At the lower boundary ($r=1 R_\odot$), the magnetic flux function
$\psi$ is fixed to the values given by the above equation, the
plasma temperature is set to be constant at 2 MK. The density,
which is assumed to be dependent on the latitude, is 10$^8$
cm$^{-3}$ and 1.5$\times 10^8$ cm$^{-3}$ in the region $\theta=[0,
30^{\circ}]$ and $[60^{\circ}, 90^{\circ}]$, respectively, and
increases linearly from $\theta=30^{\circ}$ to $60^{\circ}$. Note
that although the assumption of an inhomogeneous density
distribution at the lower boundary results in a discontinuous
latitudinal pressure gradient there, essentially similar physical
processes occur when assuming a homogeneous density distribution,
as revealed by calculations not presented here. The radial and
latitudinal velocity components at the lower boundary are obtained
by the mass conservation and the parallel condition given by
$\emph{\textbf{V}}\parallel \emph{\textbf{B}}$. The azimuthal
components of the velocity and magnetic field are set to be zero.
All parameters are extrapolated to get the conditions at the top
boundary, and symmetric conditions are employed at both the
equator and the pole. We are aware that the boundary conditions
are somewhat arbitrarily chosen in order to obtain a solution with
reasonable streamer structures and solar wind parameters, they are
still in line with available observational constraints. The
initial solar wind condition is an educated guess which does not
have impacts on the final large scale converged numerical solution
as confirmed by the calculations. For example, the radial velocity
is set to be a function of radial distance increasing
monotonically from 1.1 km/s at the bottom to 400 km/s at the top.
The contour map of the two-dimensional distribution of the
prescribed radial velocity field is depicted in Figure 1a. With
the initial and boundary conditions described above, the coupling
between the plasma dynamical motion and the coronal magnetic field
is self-consistently taken care of using our numerical code. The
calculation finally converges to a quasi-steady solution featured
by a triple-streamer, slow solar winds emanating from the poles
and between the closed arcades, and plasma blobs continuously
released from the streamer tips, as will be depicted in the
following section.

\section{The triple-streamer solar wind-blob system}

To save computation time required for the solution to converge,
the calculations were first conducted in the same domain but with
much less grid points in the latitudinal direction than described
in the last section till a quasi-steady state was reached, which
was redistributed to the present denser grid points. In the case
with lower resolution the number of grid points in the latitudinal
direction is taken to be 150 with grid spacing uniformly
distributed between $\theta=90^{\circ}$ to $\theta=30^{\circ}$
with a resolution of 0.5$^{\circ}$, and increasing according to a
geometric series of a common ratio of 1.046 from
$\theta=30^{\circ}$ to the pole. The number and distribution of
grid points in the radial direction are taken to be the same in
all calculations. The obtained solution was then allowed to evolve
self-consistently for a long time, e.g., for a few hundred hours.
The magnetic field lines with contour map of the radial velocity
field of a solution at an arbitrarily-selected instant are shown
in Figure 1b. For the purpose of illustration, the time of this
solution is set to be zero (i.e., t=0). As mentioned previously
and clearly indicated by Figures 1a and 1b, the original
quadrupolar topology of the magnetic field is reshaped
significantly by the presence of the solar wind plasmas. The most
apparent change is the formation of the triple-streamer system
with solar wind plasmas streaming into interplanetary space from
both polar regions and the narrow regions between the central and
the side streamers. In the outer part of the original side arcade
system a few field lines are opened by the solar wind plasmas,
while a few other field lines reconnect at the current sheet along
the equator and close back resulting in an increase of the closed
magnetic flux of the central arcades. This causes the
disappearance of the original neutral line at the intersection of
the red curves ($\psi=0.3697 \psi_0$) in Figure 1a, and has a
profound impact on the break-out model for CMEs proposed by
Antiochos et al. (1999) based on the field topology shown in
Figure 1a. We will return to this point in the last section. The
field lines with the same value of $\psi$ are also plotted in
Figure 1b to 1d for comparisons.

In Figures 1c and 1d we illustrate the magnetic field topology and
the radial velocity field at t=21.5 and 30 hours, respectively, so
as to indicate the long-term evolution of the dynamical streamers.
To show features in the contour map of the radial velocity, the
magnetic field lines are omitted in one of the hemispheres in
Figure 1b to 1d. An accompanying mpeg animation of the temporal
evolution of the solution from t=0 to 48 hours with frames
separated by 10 minutes in time interval is provided with the
online version of this paper. The solution at t=21.5 hours will be
re-shown in the following figure to reveal the streamer dynamics
in more detail. Note that the equator is taken as a boundary with
symmetric conditions, we are aware that such treatment certainly
has impacts on the dynamics associated with the central streamer.
We will revisit this issue again as we proceed. Therefore, in this
article we concentrate only on the dynamics atop the side
streamer. Yet, the modelling results in the central streamer
region can still be used for the purpose of comparison.

It is obvious from Figure 1 and the online animation that the
streamer is in a constant state of motion with small
magnetic/plasma islands released continuously from the elongated
tips of the side streamers, despite the long-term large-scale
stability of the overall morphology. As indicated by the
solutions, generally speaking, the islands have no regular shapes,
the latitudinal width is about 0.1 to 0.2 $R_\odot$, and the
radial length about 1 to 1.5 $R_\odot$ when the islands are first
separated from the main streamer. The plasma inside the islands is
denser than the surrounding solar wind plasmas. For example, for
the solution at t=23.5 hours, which will be further illustrated in
the following figures, the center of an island just released is
located at about 3.6 $R_\odot$. The density at this distance is
about 1.5 $\times 10^6$ cm$^{-3}$ inside the island, and 1.4
$\times 10^6$ cm$^{-3}$ in the nearby equatorward solar wind and
about 1.0 $\times 10^6$ cm$^{-3}$ in the nearby polarward wind.
The quantities of our modelled magnetic islands are basically
consistent with the LASCO observations for streamer blobs reported
by Sheeley at al. (1997) and Wang et al. (1998), and summarized in
the introduction part of this article. We therefore regard them as
streamer blobs hereinafter. The white field line delineating the
outer boundary of streamer blobs in Figure 1 and the online
animation is given by $\psi=0.65 \psi_0$. To reveal more details
of the blob dynamics, we also plot three more adjacent magnetic
field lines with $\psi=0.655, 0.66, 0.665 \psi_0$ in green, red,
and yellow colors, referred to as line A, B, and C, respectively.
It can be seen that among them line C shows no apparent motion
during the process, indicating that the blob formation involves
only the localized cusp and current sheet region. The process does
not result in the loss of the closed magnetic flux of the side
streamer.

As seen from the online animation, daughter blobs are found to
form as a result of the breakup of a longer mother blob. After the
formation, blobs change their length and shape continuously during
the outward propagation. They usually get wider as a result of the
relaxation of magnetic tension of reconnected field lines. It is
also observed that two approaching blobs with differential speeds
may get merged into one blob structure at certain distances. After
checking the animated solution carefully, we get an average
occurrence rate of about 6-8 blobs per day. Note that this
estimate has excluded the blobs that are possibly too small to be
measurable with current coronagraphs. The propagation velocity
profiles of blobs, as estimated from our numerical results, are
found to basically follow that of the surrounding solar wind
plasmas. These modelling features can be regarded as in a
remarkable agreement with available observations at the quiet time
of solar activity considering the simplicity of our model.

As clearly seen from the velocity maps depicted in Figure 1b to
1d, there exists apparent velocity shear on the two sides of the
blob forming region. It is natural to surmise that the formation
of blobs is related to the presence of the strong velocity shears
that are capable of driving various modes of streaming MHD
instabilities. To get deeper insights of how blobs are formed, in
Figure 2 we show 10 image frames spanning from t=19 to t= 23.5
hours, separated by 30 minutes in time interval with time
increasing from left to right and top to bottom. All images are
the side streamer part cropped from their corresponding global
versions which are similar as that shown in Figure 1. For example,
Figure 2f is corresponding to Figure 1c (t=21.5 hours). From
Figure 2a, we see that a narrow long blob (referred to as blob I)
is just formed from magnetic reconnection and starts its outward
propagation. About 4 to 4.5 hours later, another blob is released
from the same streamer (referred to as blob II) following the
trend of its predecessor. Therefore, the images collected in
Figure 2 delineate a complete cycle of blob formation of typical
configuration. Below we shall describe the evolutionary process
and physical interpretation of the formation of blob II by
illustrations with Figures 2 and 3.

For a visual guide, in Figure 2 we add three black arcs centered
at the sun with radius being 2.35, 2.6, and 3.6 $R_\odot$. We see
from Figure 2a that the streamer cusp is reformed at a lower
distance of about 2.6 $R_\odot$ upon the disconnection of blob I.
The magnetic field at the newly reformed cusp region is apparently
too weak to sustain the confinement of the plasmas at coronal
temperature, the plasmas expand and stretch the freezing-in closed
field lines radially outwards. Consequently the streamer tip
elongates as clearly seen from the upper panels of Figure 2. After
a continuing expansion for about 2 to 2.5 hours line A (in white
color with $\psi=0.655 \psi_0$) starts to pinch at its middle
part. It is apparent that the physics accounting for the pinching
of field lines drives the following reconnection shown in Figure
2i and 2g. To check the physical conditions at the initial stage
of the line pinch, in the left panel of Figure 3 we plot the
radial components of velocity and magnetic field vectors ($v_r$
and $B_r$) together with the Alfv\'en speed ($v_A$) for the
solution shown in Figure 2f (t=21.5 hours) as a function of
co-latitude ($\lambda=90{^\circ}-\theta$). The profiles are
plotted at the 3 heliocentric distances used for the visual guide,
i.e., 2.35 (solid), 2.6 (dotted), and 3.6 (dashed) $R_\odot$.

From the $v_r$ profile, we see that there exists a strong velocity
jump between the surrounding solar wind plasmas along open field
lines and that inside the elongated closed loops reaching up to
nearly 150 km s$^{-1}$ at 2 to 3 solar radii. In the shear region
the magnetic field changes its direction, and the overall strength
is relatively weak compared to that in the surrounding solar wind
regimes. Note that $B_r$ at 2.6 $R_\odot$ (the dotted line)
changes its polarity 3 times in the shear region. This feature is
an aftermath of blob I, and will be interpreted later together
with the physical origin of the small concave morphology of lines
A and B at the streamer tips upon the formation of blobs I and II.
The Alfv\'en speed reaches the minimum value inside the velocity
shear, as a result of the weak magnetic field there. Generally
speaking, from 2.35 to 3.6 $R_\odot$ the Alfv\'en speed is less
than 75 km s$^{-1}$ in the solar wind flowing between the central
and the side streamers, much higher in the polar wind regime, and
less than 50 km s$^{-1}$ inside the velocity shear region.

Velocity shears in plasmas can generate Kelvin-Helmholtz (KH)
instabilities (e.g. Miura \& Pritchett 1982). The physical
configuration as shown in Figure 3a to 3c is somewhat similar to
that studied previously by, e.g., Lee, Wang {\&} Wei (1988) for
streaming instabilities (more widely known as KH instability),
although they investigated the situation with velocity maximum
located inside the current sheet region, on the contrary to our
situation. Such discrepancy can be easily reconciled if one
applies a coordinate transformation from their configuration into
a moving reference system with an appropriate velocity, e.g., the
maximum velocity in their configuration. With the transformation,
the real part of the frequency is doppler-shifted, and the
instability criteria, eigenmode profiles, and growth rate remain
unaffected. Using their linear model of streaming instability, Lee
et al. (1988) determined the eigenmode profiles and critical
conditions (see Equations (41) and (42) in their article) for both
the sausage and the kink modes. To take advantage of the deduced
conditions, we need to specify the Alfv\'en speed in the shear
region and in the surrounding plasmas, the number density, the
wave number, and the half width of the shear region. Referring to
the parameter profiles plotted in Figure 3a to 3c, we assume an
Alfv\'en speed to be 20 km s$^{-1}$ inside and 75 km s$^{-1}$
outside the shear region, the half width of the shear to be 0.05
$R_\odot$, and the density to be distributed homogeneously. After
simple calculations with Lee et al.'s critical conditions, we find
that in the hypothetical configuration for the specified wave
length of 1 $R_\odot$ (typical length of a blob), the velocity
gradient should be no less than 158 km s$^{-1}$ for the sausage
mode and 100 km s$^{-1}$ for the kink mode to take place. By
checking Figure 3a, we find that the total velocity jump presented
in our solution at the time field line A starts to pinch can
indeed reach up to 150 to 160 km s$^{-1}$. This implies that both
the sausage and the kink modes of streaming instabilities can
occur in our configuration. We emphasize that the above estimate
is rather rough and carried out with a much simpler configuration
than that represented by Figure 3a to 3c. It is well known that
the sausage mode results in the variation of the current sheet
thickness to form a sausagelike configuration, and the kink mode
produces a wavy structure keeping the current sheet thickness
relatively constant. By checking the online animation again, it is
apparent that both the sausagelike and wavy structures are present
in the blob forming region. Thus, the occurrence of the mixed
sausage-kink mode instability in our solutions is actually
self-evident. On the other hand, the symmetric boundary conditions
adopted at the equator effectively prohibit the growth of the kink
mode. Hence only streaming sausage mode can be observed atop the
central streamer.

From the above text, we propose that the pinch of field lines is
caused by the streaming sausage mode instability. The instability,
ideal in nature, is unable to produce reconnections. Yet, with the
nonlinear growth of the instability, the distance decreases and
the current density increases consistently in between the
elongating field lines till whatever dissipation mechanism sets in
to cause reconnection. Therefore, we suggest that the
reconnections accounting for the blob separations are driven by
the nonlinear development of the sausage mode. The reconnections
could be caused by the well-known resistive tearing mode
instability, or simple magnetic diffusion process across the
narrow layer with enhanced current density. To determine the exact
process acting in this highly-nonlinear MHD simulation, one needs
to consider a more sophisticated and physical model of resistivity
and a finer numerical resolution to reduce the numerical
resistivity as low as possible. This is generally rather robust
and computationally expensive, and remains unresolved in the
current study. As apparently seen from the multi-blob formation in
our numerical solution, reconnections can take place
simultaneously at various locations. The wavelength of the sausage
mode is reflected by the length of the blob just formed, which is
about 1.0 $R_\odot$ to 1.5 $R_\odot$. The upper limit of the
growth time of the sausage mode can be estimated by measuring the
time span from a field line starts to pinch to a blob is finally
released, which is about 2 to 3 hours according to Figure 2e to
2j.

The sausage mode instability is energized by the free energy
associated with the velocity shear present in the blob forming
region. To see how velocity shear evolves after the formation of a
blob, in the right panels of Figure 3 (3d - 3f) we plot the
latitudinal profiles for the solution shown in Figure 2j (t=23.5
hours) at the same three distances as used in the left panels.
From Figure 2j, it can be seen that the break point of the
streamer is located at about 2.6 $R_\odot$, the profiles at this
distance are plotted as dotted lines. The velocity shear almost
vanishes crossing the streamer break point, as the aftermath of
the sausage-mode instability. It can be seen from the solid lines
plotted in both Figures 3a and 3d, strong velocity shear is always
present at a lower distance of 2.35 $R_\odot$. Such velocity shear
is along the streamer leg between the open and closed magnetic
field lines, where the local magnetic field is relatively strong
and a KH instability is unable to develop. In the following 2 to 3
hours of continuous streamer elongation after t=23.5 hours, the
strong velocity shear is brought into the slow wind regime again
where the field strength gets lower so that the critical
conditions for streaming instabilities can be satisfied.

Similar to the profile for 2.6 $R_\odot$ plotted in Figure 3b, the
radial magnetic field at about 2.35 $R_\odot$ in Figure 3e also
presents a triple change of polarity. This feature can be
straightforwardly explained by the concave-inward morphology of
line B and its adjacent lines not plotted here. The formation of
the concave-inward morphology can be interpreted as follows. The
causes are different for different field lines. For the field line
that is just broken by reconnections, like line B (in red color)
in Figure 2i, the morphology is a result of the relaxation of
intensified magnetic tension of the over-stretched field lines
upon reconnections. For the field line overlying the lines just
reconnected, like line A in Figures 2a and 2i, the morphology is a
result of magnetic pressure gradient force which points inwards as
a result of the rapid shrinking of those lines just reconnected
(like line B).

As a summary of this section, the formation of streamer blobs can
be regarded as consisting of two successive processes, one is the
plasma-magnetic field expansion through the localized cusp region
where the field is too weak to sustain plasma confinement, the
other is the onset of the sausage mode instability that finally
triggers magnetic reconnections producing separated plasma blobs.
The former process brings strong velocity shear into the slow wind
regime, and provides required free energy for the following
streaming instabilities. By checking the time frames collected in
Figure 2 for a complete cycle of the formation of a typical blob,
we find that generally the streaming sausage mode instability sets
in after the expansion continues for about 2 to 2.5 hours. The
expansion continues till the nonlinear development of the
instability finally breaks the elongated loops, which takes
another 2 to 3 hours, producing a streamer blob and a {`}new'
coronal streamer with a lower cusp. The whole process originates
from the peculiar magnetic topological feature at the cusp region
intrinsically associated with a coronal streamer. We therefore
term the described process as intrinsic instability of coronal
streamers. Although the instability is intrinsic, it does not lead
to the loss of the closed magnetic flux, so it does not influence
the overall feature of a streamer.

\section{Conclusions and discussion}

It is well known that a coronal streamer system is resulted from
the dynamical equilibrium between plasma expansion and magnetic
confinement. Despite the overall long-term stability of the
morphology, it is suggested that the streamer is subject to an
intrinsic instability originating from the peculiar magnetic
topological feature at the cusp region. The whole process of the
instability consists of two successive processes, one is the
plasma-magnetic field expansion through the localized cusp region
where the field is too weak to maintain plasma confinement, the
continuing expansion brings strong velocity shear into the slow
wind regime providing free energy necessary for the onset of a
streaming sausage mode instability which causes pinches of
magnetic field lines and drives reconnections at the pinching
points to form separated magnetic blobs.

Although our recent two-dimensional models for the corona and
solar wind have considered the role of Alfv\'en waves in the
formation of the bipolar coronal streamer and the solar wind as a
two-fluid and two-state plasmas (Chen \& Hu, 2001; Hu et al.,
2003a), in this paper we still take advantage of the polytropic
assumption of the heat transport process for simplicity. These
previous streamer and solar wind models were aimed at
investigating the heating and acceleration of the corona and solar
wind plasmas emphasizing the large-scale stability of the corona
streamers rather than the detailed dynamics of blobs. To do this,
a certain amount of numerical dissipation was applied to stabilize
the calculations. As a drawback of this treatment, the dynamical
processes associated with blob formations have been largely
smoothed out in these previous models. Note that although no
explicit heating is applied to the plasma in the present model, a
polytropic index of $\gamma=1.05$ and a temperature of 2 MK at the
inner boundary results in the high-temperature coronal plasma with
a nature to expand. In a future extension to the present model,
one may let $\gamma={5 \over 3}$ and employ a more realistic
energy balance equation to model the plasma heating process. This
will change the details of the plasma-magnetic coupling process,
and thus the streamer morphology as well as the solar wind and
blob properties. However, we believe that the overall physics of
blob formation suggested above, namely, the instabilities are
essentially due to the presence of an intrinsic streamer cusp with
the dynamics determined by the elongation of the freezing-in
plasma-magnetic arcades, which brings the relatively slow confined
plasmas into the solar wind regime resulting in enough velocity
shear to trigger the streaming sausage mode instability, remains
largely unaffected.

In the study by Lapenta \& Knoll (2005), blobs are interpreted as
a result of reconnections along open current sheet structure
beyond the streamer cusps. This can be seen by comparing their
Figure 6a and Figure 6d (also 7a and 7c, 7d and 7f). At the end of
their case calculation there is an apparent increase of the
closed-flux inside the streamer. The increase can be more than
50-100{\%} of the original total closed flux contained by their
streamer if the magnetic flux function along adjacent field lines
plotted in their figures is equally separated. Thus, the streamer
itself will grow significantly and continuously after blobs form
one by one. Although the magnetic field is not directly observable
in the corona, this shall be inconsistent with observations which
indicate that blobs have only minor contributions to the solar
wind plasmas resulting in no dislocation or disruption of the
streamer. In our study, blobs are thought to form along elongating
closed arcades below the streamer cusp. The closed streamer flux
is not influenced by the blob formation in any significant way.
Furthermore, according to our calculations, reconnections occur in
the region where the field lines are almost parallel to each
other, rather than in the region where flows/field lines converge
most, therefore, the idea of converging-flow driven reconnection
to form blobs, as proposed by Lapenta \& Knoll, seems to be
inappropriate at least in our more realistic streamer
configurations.

Our calculations started from the quadrupolar magnetic field with
a neutral point high in the corona, exactly in the form provided
by Antiochos et al. (1999) in their break out model for CMEs. They
investigated the dynamical response of the closed arcades to foot
point shearing motions. One key point of the breakout model lies
in the presence of a multipolar pre-eruptive field with a magnetic
null (or neutral point) high in the corona. With the shearing
motions, the high magnetic null gradually extends into a current
sheet structure that manages to survive for a certain time in the
corona plasmas. Through magnetic forces the current sheet
structure serves as a major agent confining the increasing free
magnetic energy of the sheared central arcades. Eruptions take
place when the thickness of the current sheet structure eventually
decreases to trigger the fast break-out reconnections through
either physical or numerical resistivity. The original breakout
model did not take into account the effect of expanding plasmas on
the global magnetic field morphology. Such effect is
non-negligible especially at the heights where the original
breakout model is concerned and at the null neighborhood with very
weak magnetic field strength. The consequence of this effect has
been clearly illustrated by the calculations presented in this
article, especially, by comparisons of the upper panels of our
first figure. We found that the original morphology of the
quadrupolar magnetic field is re-shaped into a triple-streamer
configuration with its neutral point blown away by the solar wind,
indicating the critical impact of the solar wind plasmas on the
coronal field topology that causes the failure of the breakout
mechanism within the framework of our modelling. In future
studies, we are going to construct a CME model based on the
triple-streamer solar wind state obtained in this paper using the
ideal catastrophe mechanism of magnetic flux rope system as the
triggering and major energizing mechanism for eruptions (see,
e.g., Chen et al., 2006, 2007).

\acknowledgements We thank Prof. Y. Q. Hu for valuable
suggestions. This work was supported by grants NNSFC 40774094,
40825014, and NSBRSF G2006CB806304 in China. Xing Li is supported
by a STFC Rolling Grant to Aberystwyth University.

\clearpage
\begin{figure}\epsscale{1.} \plotone{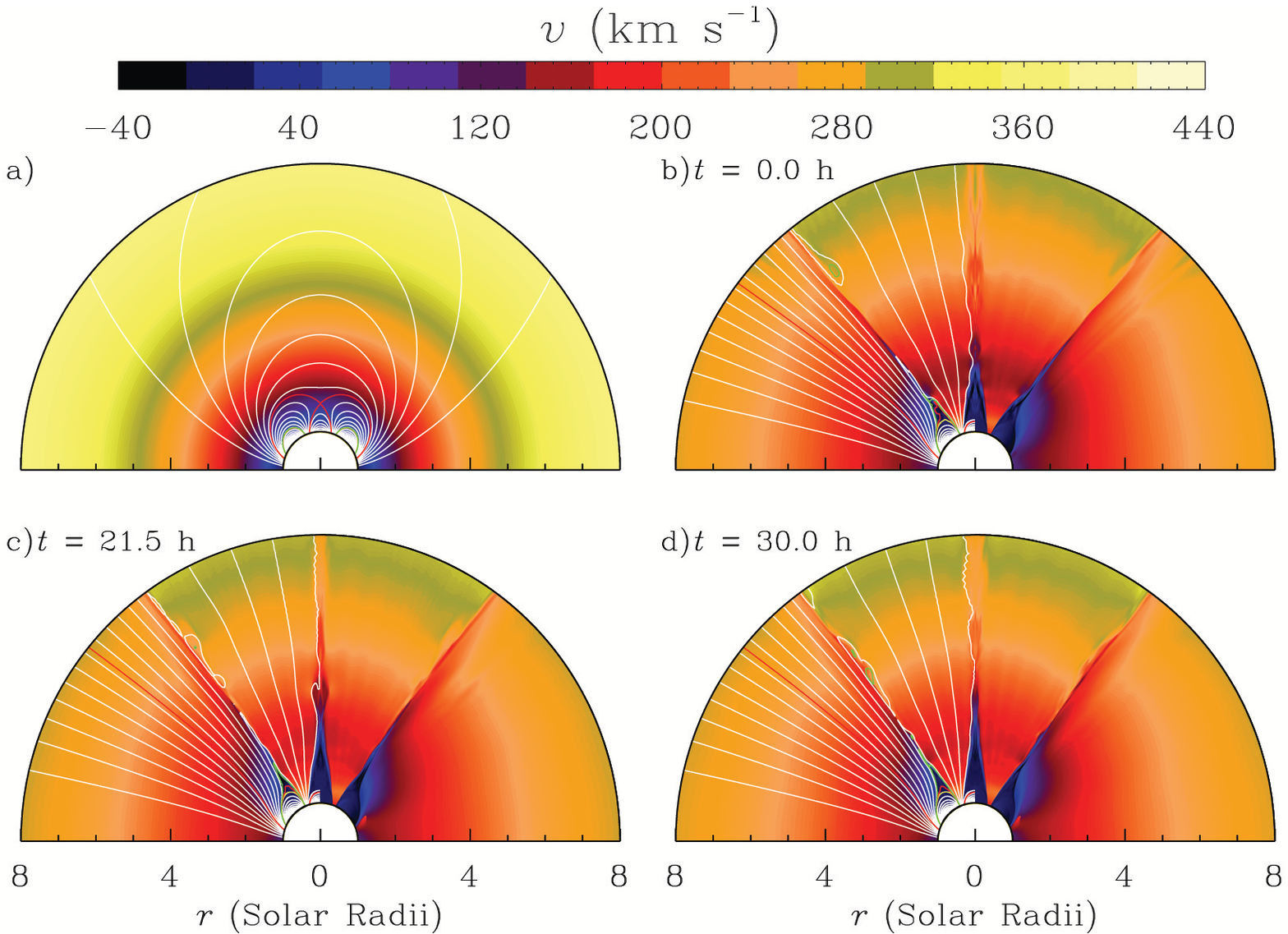}
\caption{Magnetic field lines superposed by the contour map of the
two-dimensional distribution of the radial velocity field from 1
to 10 $R_\odot$. The field lines are represented by the contours
of the magnetic flux function with the contour level separated by
$\psi=0.05 \psi_0$. Panel a) is for the initial quadrupolar field
and solar wind velocity, b) - d) depict the calculated conditions
at various instants. The separatrix associated with the original
quadrupolar field ($\psi=0.3697 \psi_0$) is shown in red curves.
The three lines in green, red, and yellow colors in the blob
region are given by $\psi=0.655, 0.66$, and $ 0.665 \psi_0$,
referred to as line A, B, and C in the text. To show features in
the contour map of the radial velocity, the magnetic field lines
are omitted in one of the hemispheres in b) - d). [\emph{The
long-term dynamical evolution (in 2 days from t=0 to 48 hours) of
the triple-streamer solar wind blob system is shown in an mpeg
animation in the electronic edition of the Journal.}]
\label{fig1}}
\end{figure}

\begin{figure}
\epsscale{1.}  \plotone{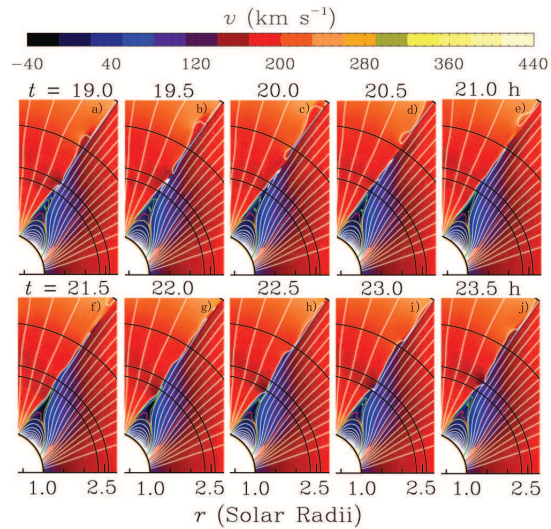} \caption{Ten image frames of the
side streamer part cropped from the corresponding global versions
of images, spanning from t=19 to t= 23.5 hours. The frames are
separated by 30 minutes in time interval with time increasing from
left to right and top to bottom. The three black arcs are centered
at the sun with radius being 2.35, 2.6, and 3.6 $R_\odot$.
[\emph{This figure is available as part of the mpeg animation in
the electronic edition of the Journal.}] \label{fig2}}
\end{figure}

\begin{figure}
\epsscale{1.}  \plotone{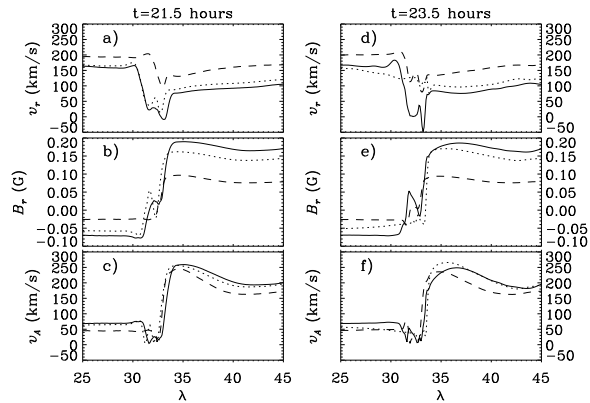} \caption{The radial components of
velocity and magnetic field vectors ($v_r$ and $B_r$) together
with the Alfv\'en speed ($v_A$) for the solutions at t = 21.5
hours left) and 23.5 hours (right). The parameters are shown as a
function of co-latitude ($\lambda=90^{\circ}-\theta$) for 3
heliocentric distances of 2.35 (solid), 2.6 (dotted), and 3.6
(dashed) $R_\odot$. \label{fig3}}
\end{figure}

\end{document}